\documentclass[12pt]{iopart}
\usepackage{natbib}
\usepackage{graphicx}
\expandafter\let\csname equation*\endcsname\relax
\expandafter\let\csname endequation*\endcsname\relax
\usepackage{amsmath}

\begin{document}

\title[MHD simulation of a flux rope in quadrupolar configuration]{Magnetohydrodynamics simulation of magnetic flux rope formation in a quadrupolar magnetic field configuration}

\author{Sanjay Kumar$^1$, Avijeet Prasad$^{2,3}$, 
Sushree S. Nayak$^4$, Satyam Agarwal$^{5,6}$, and R. Bhattacharyya$^5$}

\address{$^1$ Department of Physics, Patna University, Patna-80005, India.}
\address{$^2$ Rosseland Centre for Solar Physics, University of Oslo, Postboks 1029 Blindern, 0315 Oslo, Norway.}
\address{$^3$ Institute of Theoretical Astrophysics, University of Oslo, Postboks 1029 Blindern, 0315 Oslo, Norway. }
\address{$^4$Center for Space Plasma \& Aeronomic Research,
The University of Alabama in Huntsville,
Huntsville, Alabama 35899, USA. }
\address{$^5$Udaipur Solar Observatory, Physical Research Laboratory, Dewali, Badi Road, Udaipur-313001, India. }
\address{$^6$Discipline of Physics, Indian Institute of Technology, Gandhinagar 382355, India.
}

\ead{sainisanjay35@gmail.com}

\vspace{10pt}
\begin{indented}
\item[] 03 November 2022
\end{indented}

\begin{abstract}
Magnetic flux ropes (MFRs) play an
important role in high-energetic events like solar flares and coronal mass ejections in the solar atmosphere. Importantly, solar
observations suggest an association of some flaring events with quadrupolar
magnetic configurations. However, the formation and subsequent evolution
of MFRs in such magnetic configurations still need to be fully understood.
In this paper, we present idealized magnetohydrodynamics (MHD)
simulations of MFR formation in a quadrupolar magnetic configuration.
A suitable initial magnetic field having a quadrupolar configuration is
constructed by modifying a three-dimensional (3D) linear force-free magnetic
field. The initial magnetic field contains neutral lines, which consist of X-type null points. The simulated dynamics initially demonstrate the oppositely
directed magnetic field lines located across the polarity inversion lines
(PILs) moving towards each other, resulting in magnetic reconnections.
Due to these reconnections, four highly twisted MFRs form over
the PILs. With time, the foot points of the MFRs move towards the X-type
neutral lines and reconnect, generating complex magnetic structures around the neutral lines, thus making the MFR topology more complex in the quadrupolar configuration than those formed in bipolar loop systems.
Further evolution reveals the non-uniform rise of the MFRs. 
Importantly, the simulations indicate that the pre-existing X-type null points in magnetic configurations can be crucial to the evolution of the MFRs and may lead to the observed brightenings during the onset of some flaring events in the quadrupolar configurations.
\end{abstract}

%
%Uncomment for keywords
\vspace{2pc}
\noindent{\it Keywords}: Magnetohydrodynamics, Magnetic reconnections, Magnetic Flux ropes, EULAG-MHD.
%
% Uncomment for Submitted to journal title message
%\submitto{\JPA}
%
% Uncomment if a separate title page is required
%\maketitle
% 
% For two-column output uncomment the next line and choose [10pt] rather than [12pt] in the \documentclass declaration
%\ioptwocol
%

\section{Introduction}
\label{intro}
 A magnetic flux rope (MFR) is a bundle of twisted magnetic field lines (MFLs) which wind around a given axis {\citep{priest2014book}}. MFRs play a vital role in the dynamical
evolution of diverse plasma systems, such as the laboratory, space, and astrophysical plasmas \citep{aschwanden2004book, moser2012, van2012, 2022AIPA...12a5110E}. Particularly, in the solar context, MFRs are believed to be the magnetic structures that can lead to eruptive events in the solar atmosphere, such as solar flares, coronal mass ejections (CMEs), and prominence eruptions {\citep{aschwanden2004book, priest2014book}}. The twisted field lines of MFRs represent regions of strong currents and facilitate the storage of magnetic energy {\citep{2011LRSP....8....6S, priest2014book}}.  These eruptive events release the stored magnetic energy in the form of radiation, mass flow, and accelerated charged particles. Thus, it is crucial to investigate the formation and subsequent evolution
of MFRs to gain a comprehensive understanding of solar eruptions. 
 
There are broadly two mechanisms through which an MFR can appear in the solar atmosphere. In the first mechanism, an MFR pre-exists below the solar surface and, due to the magnetic buoyancy, can become unstable and emerge out of the solar surface into the atmosphere {
\citep{2006ApJ...637L..65G,  2009LRSP....6....4F,  Fan_2010, 2015ApJ...806...79F}}. In contrast, the second mechanism advocates the formation of MFR in the solar atmosphere itself by magnetic reconnections (MRs) inside sheared magnetic arcades {\citep{1989ApJ...343..971V, 2001ApJ...552..833M, amari+2003apj, aulanier2010}}. These reconnection-based studies of MFR formation often rely on the force-free field assumption (i.e., vanishing Lorentz force) of the coronal field. 
The initiation of MRs and subsequent MFR formation are governed by prescribed (shearing and/or converging) flows at the bottom boundary. Recently, an alternative approach based on non-force-free fields is also developed that supports the non-zero Lorentz force  {\citep{sanjay2016, prasad+2017apj, prasad+2018apj,  prasad_2020}}. This approach demonstrates the spontaneous onset of MRs in arcades without any prescribed boundary flow, accounting for the development and subsequent evolution of MFRs.

It is noteworthy that the MFR, once formed, can exist as a stable structure and manifest itself in the form of filament/prominence {\citep{aschwanden2004book}}. With time, they can become unstable by ideal magnetohydrodynamics (MHD) instabilities such as the kink instability or the torus instability {\citep{2005ApJ...630L..97T, kliem&torok2006prl}} --- leading to eruptions. Alternatively, the MFR can remain in the dynamic phase due to repeated MRs, which further add a significant amount of magnetic fluxes in the MFR --- causing a rise and expansion of the MFR {\citep{sanjay2016, prasad+2017apj}}. Such rapidly rising MFRs can play a crucial role in the onset of a solar flare and the subsequent evolution of the eruptions 
{\citep{2011LRSP....8....6S, prasad+2017apj, prasad_2020}}.  The numerical studies on MFR formation through MRs so far have been predominantly carried out for the bipolar flux systems {\citep{1989ApJ...343..971V, amari+2003apj, aulanier2010, sanjay2016,  prasad_2020}}, while far more complex topologies are observed in the extreme-ultraviolet (EUV) observations of the solar corona. Particularly, some observational studies have revealed flaring regions to be associated with quadrupolar magnetic configurations \citep{1997ApJ...489..976N, 2012ApJ...757..149S, 2017ApJ...842..106K, 2017ApJ...843...93C, 2022ApJ...926..143M}. 
Although quadrupolar configurations have been employed in previous works \citep{ 1999SoPh..186..301H, 2012SoPh..277..131R} to assess the magnetic topology and energy storage in the coronal field, not many studies have been conducted to explore the formation and evolution of MFRs in such configurations. 
Using a flux emergence-based model, \citet{2015ApJ...806...79F} has shown the formation of the complex quadrupolar magnetic structures in the solar corona, whereas \citet{2017ApJ...846..106L} used a magnetofrictional model for automated detection of magnetic flux ropes and studying MFR formation in the solar corona.
Another recent study by {\citet{2019A&A...630A.134S}} examined the evolution of MFRs in quadrupolar configuration and the role of MFR evolution in solar flares utilizing pre-existing twisted magnetic structures emerging from below the photosphere.

%Using the MHD simulations, {\citept{2019A&A...630A.134S}} has attempted to explore the evolution of MFRs in quadrupolar magnetic configurations and the role of the evolution in the flaring events. Notably, these simulations judiciously preconditioned the MFR evolution by specifying preexisting twisted magnetic structures that emerge from below the photosphere. 
%In contrast, in this work, we present idealized MHD simulations of the evolution of MFRs in terms of their formation from initial quadrupolar magnetic loops and continuous ascent, mediated via magnetic reconnections.}}
%Moreover, in previous works \citep{ 1999SoPh..186..301H, 2012SoPh..277..131R},  the quadrupolar configurations are used to assess the magnetic topology and energy storage for the coronal field using different force-free models. However, the MFR formation and its evolution in quardupolar configurations are not explored 
%Relevantly, recent studies {\citep{2017ApJ...842..106K, 2022ApJ...926..143M}} have  revealed the flaring regions associated with the quadrupolar magnetic configurations. Moreover, \citept{2022ApJ...926..143M} have indicated the role of MFRs in an M-class flare from a quadrupolar magnetic configuration. 

In this work, we explore MFR evolution in terms of their formation from quadrupolar magnetic loops and continuous ascent, mediated via magnetic reconnections. We simulate the viscous relaxation of an
incompressible, thermally inactive, and infinitely conducting plasma from an initial non-equilibrium state to ensure the spontaneous onset of MRs in the MHD simulation {\citep{kumar+2014phpl, sanjay2016}}. 
The spontaneity of MRs stems from Parker’s magnetostatic theorem {\citep{parker1972, parker1994}},
according to which a plasma with infinite electrical
conductivity and complex magnetic topology can not achieve an equilibrium state with a continuous magnetic field. The viscous relaxation is in
harmony with the magnetostatic theorem --- leading to the natural generation of MRs {\citep{kumar+2014phpl, sanjay2016}}. The initial non-equilibrium state for the relaxation is achieved through the initial non-force-free field, which provides the Lorentz force that initiates dynamics without needing a prescribed flow. Relevantly, a recent study by \citet{2022SoPh..297...91A} has suggested the similarity in the magnetohydrodynamics of a coronal transient initiated with force-free and non-force-free extrapolated magnetic fields. Furthermore, in the context of the relaxation theory, the minimum-dissipation-rate (MDR) principle results in non-force-free fields as a relaxed state for externally driven systems {\citep{bhattacharyya+2007soph, 2010JPlPh..76..107S}}. This makes the solar corona, which is driven by the photospheric boundary flows, a suitable candidate for the application of the MDR principle {\citep{bhattacharyya+2007soph}}. The physical scales can become under-resolved during the relaxation constrained by the flux-freezing condition. Our numerical scheme intermittently and adaptively regularises these  under-resolved scales by mimicking magnetic reconnections. The simulated evolution documents the onset of magnetic reconnections, which lead to the formation of the MFRs in the quadrupolar magnetic configuration. The MFRs are found to be morphologically more complex than the ones generated in the bipolar magnetic loops. Subsequently, the reconnections continue in time and add more magnetic flux to the MFRs, causing the rise of the MFRs.   

The paper is organized as follows. The initial magnetic field is described in Section 2. The governing MHD equations and the numerical model are discussed in Section 3. The simulation results are presented in Section 4. Section 5 summarises these results and discusses the key findings.

\section{Initial Magnetic Field}
\begin{figure}[htp]
\centering
\includegraphics[width=0.9\textwidth]{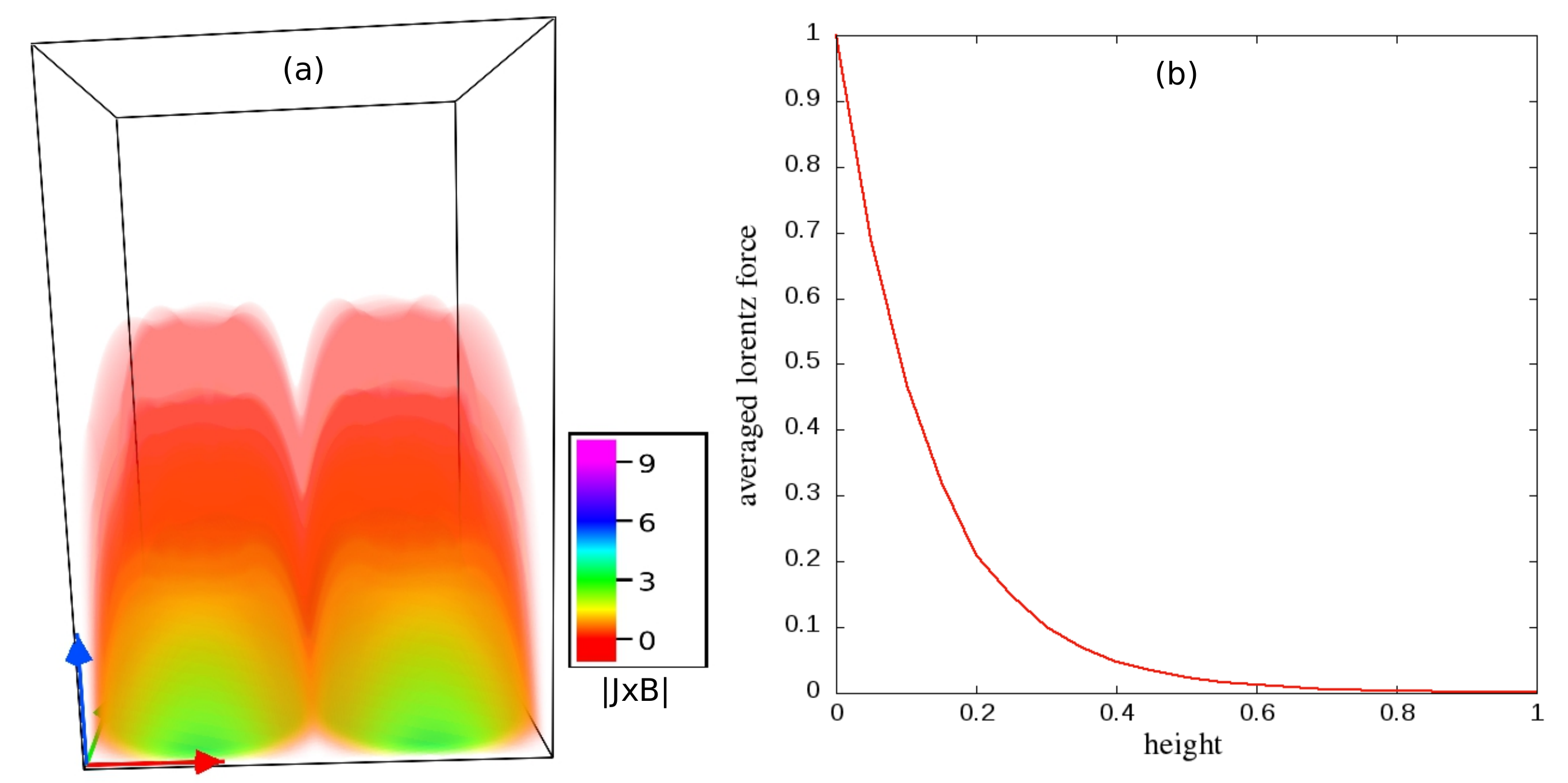}
\caption{Panel (a) illustrates the distribution of the magnitude of the initial Lorentz force density in the computational domain. Panel (b) plots the magnitude variation of the horizontally averaged Lorentz force density with height. The force density and height are normalized to their
maximum values, as our focus is on its decay rate with height. Both panels clearly illustrate the exponential decrease of the force density with
height. } \label{figure1a}
\end{figure}

To achieve a suitable initial magnetic field with quadrupolar configuration, we revise the general construction procedure utilized by \citet{sanjay2016}, which was aimed to study the MFR formation in the bipolar magnetic loops. The Cartesian components of the chosen initial magnetic field $\bf{B}$ are

%\begin{equation}
\begin{align}
%\begin{split}
{B_{x}}&=\quad\! 0.5\bigg[ \alpha_0 \sin\left( x\right) \cos \left(y
\right) \exp\left( -\frac{k_0z}{s_0}\right) - k_0\cos\left( x\right) \sin\left( y\right) \exp\left( -\frac{k_0z}{s_0}\right) \bigg], \label{bx}\\
% \end{split}
%\end{equation}
%\begin{equation}
%\begin{split}
{B_{y}} &= -0.5\bigg[ \alpha_0 \cos\left(x\right)\sin\left(y
 \right) \exp\left( -\frac{k_0z}{s_0}\right) 
 + k_0\sin\left( x \right) \cos\left( y\right) \exp\left( -\frac{k_0z}{s_0}\right) \bigg],\label{by}\\
 %\end{split}
% \end{equation}
% \begin{equation}
%\begin{split}
{B_{z}} &= \qquad \quad \! s_0\sin\left( x \right) \sin\left( y\right) \exp\left(-\frac{k_0z}{s_0} \right),
 \label{bz}
 %\end{split}
% \end{equation}
\end{align}

\noindent where $\alpha_0$, $k_0$ and $s_0$ are constants. The constant $\alpha_0$ and $k_0$ are related as $k_0=\sqrt{2-{\alpha_0}^2}$. The field ${\bf{B}}$ is specified in the positive half-space ($z \ge 0$) of a Cartesian domain $\Gamma$, which is periodic along the lateral directions (ranging from $0$ to $2\pi$ in $x$ and $y$) and open along the vertical direction (ranging from $0$ to $6\pi$ in $z$).  The initial magnetic field ${\bf{B}}$ supports the non-zero Lorentz force for $s_0 \ne 1$, which contributes to the viscous relaxation.  For the simulations, we set $s_0=6$ to have an optimal Lorentz force for generating efficient dynamics with a minimal computational cost. Noticeably, for $s_0=1$,  ${\bf{B}}$ satisfies the linear force-free equation  $\nabla\times{\bf{B}}=\alpha_0 {\bf{B}}$, where $\alpha_0$ represents the magnetic circulation per unit flux, and is related to the twist of the corresponding MFLs \citep{kumar+2014phpl}.  We set $\alpha_0=0.1$ to have sufficiently twisted initial MFLs and, hence, complex magnetic topology (as demanded by the magnetostatic theorem).  Moreover, the chosen $\alpha_0$ corresponds to a high value of $k_0$ that leads to a steeper exponential decay of the initial Lorentz force (see Figure \ref{figure1a}). 
Relevantly, the solar corona is considered to be in the force-free equilibrium state under the low plasma-$\beta$ approximation {\citep{priest2014book}}. On the other hand, the photosphere with plasma-$\beta \sim 1$ is expected to be non-force-free due to the convective driving \citep{2001SoPh..203...71G}.

 In Figure \ref{figure1a}(a), we have shown the
direct volume renderings of the Lorentz force density, which shows the decay of the Lorentz force with height. In this and the subsequent figures, the arrows in the colors red, green, and blue represent the directions $x$, $y$, and $z$, respectively. 
To confirm this further, we plot the horizontally averaged Lorentz force density variation with height in Figure \ref{figure1a}(b). The averaged Lorentz force density falls off sharply with height, such that the upper half of the domain is nearly in a force-free state.

\begin{figure}[htp]
\centering
\includegraphics[width=0.9\textwidth]{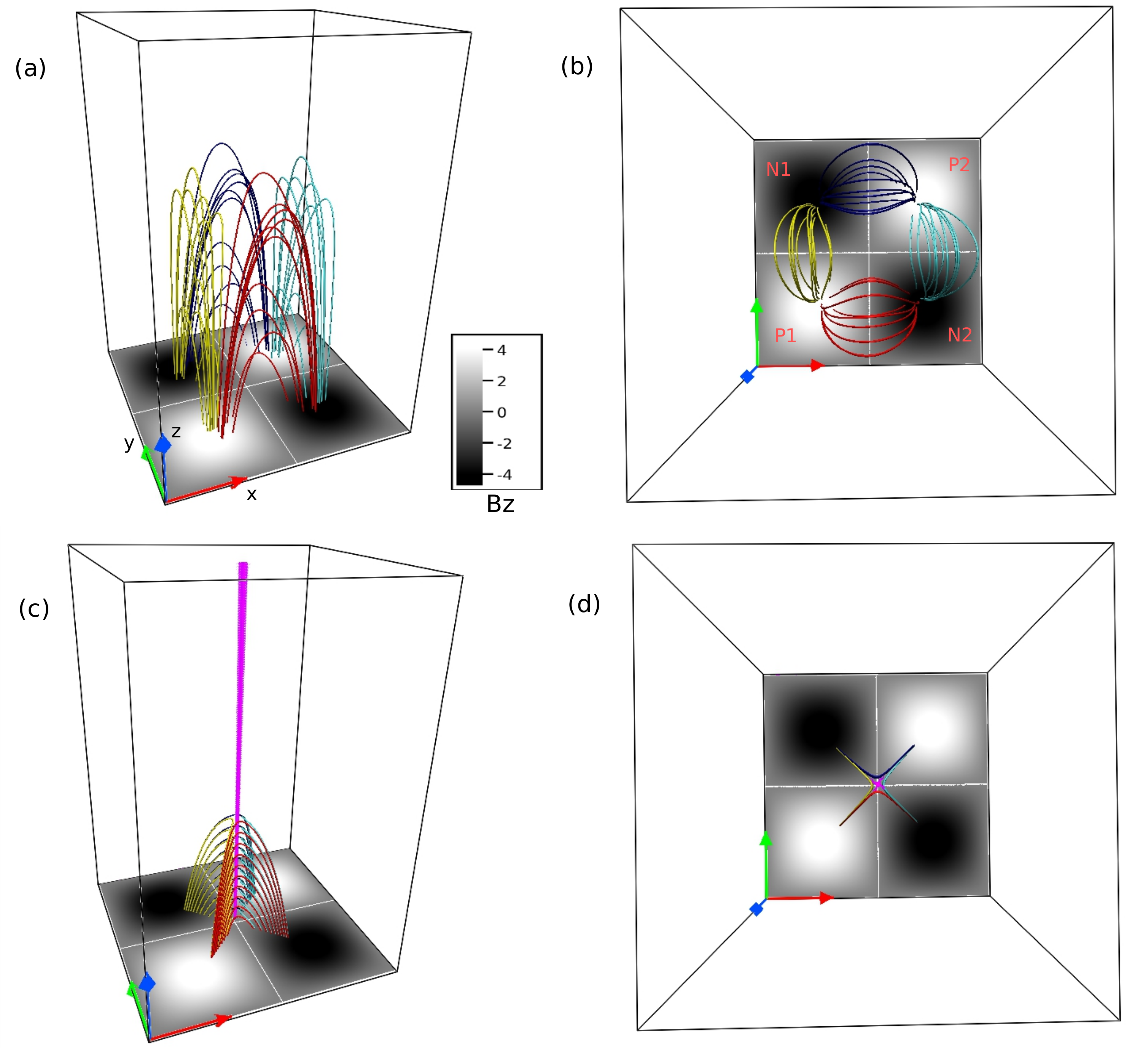}
\caption{The figure depicts the initial magnetic field overlaid with $B_z$-contours at the bottom boundary. The corresponding magnetic field lines have a geometry similar to the coronal loops (panel (a)). The quadrupolar magnetic configuration is evident in panel (b) by the presence of two positive polarities (marked by P1 and P2) and two negative polarities (denoted by N1 and N2) separated by PILs (white lines). Panel (c) shows the existence of a neutral line (in pink) at $(x, y)=(\pi, \pi)$ along the $z$ axis inside the domain. The X-type geometry of the field lines near the neutral line (see panel (d)) confirms that the line is made of X-type neutral points.} \label{figure1}
\end{figure}

Figure \ref{figure1} shows the initial magnetic field configuration. In the figure, the lower boundary is superimposed with the $B_z$ values. 
%Notably, the figure illustrates the MFLs to be geometrically similar to the coronal loops with a non-zero twist. 
The multiple polarities in the initial field are shown in Figure \ref{figure1}(b), where P1 and P2 mark the positive polarities, while N1 and N2 denote the negative polarities.
The polarity inversion lines (PILs) corresponding to the different opposite polarities (P1, N1), (P1, N2), (P2, N1), and (P2, N2) are shown by white lines in the figure. Noticeably, the opposite polarities satisfy mirror symmetry across the PILs. Because of the mirror symmetry, the initial magnetic field supports an X-type neutral line (made of magnetic neutral points with X-type field line geometry) located at ($x, y$)=($\pi, \pi$) along $z$ (shown in pink in panel (c)). 
As a result, when seen from the top, the initial field line geometry is quadrupolar (panel (d)). 
In addition, since the computational domain is periodic along $x$ and $y$, the initial field also has neutral lines at the domain's boundaries (not shown). The locations of these lines are ($x, y$)=($0, 0$), ($0, \pi$), ($0, 2\pi$), ($\pi, 0$), ($2\pi, 0$), ($\pi, 2\pi$), ($2\pi, \pi$), and ($2\pi, 2\pi$) along $z$ axis. 
The initial magnetic field with the two positive and negative polarities resembles the observed quadrupolar magnetic configurations at the solar surface {\citep{2017ApJ...842..106K, 2022ApJ...926..143M}}. However, in the absence of mirror symmetry and periodicity, the observed configurations exhibit a more complex magnetic field line topology.   

\section{Governing MHD equations and Numerical Model} 
With a focus on exploring the changes in field line topology, here we consider the plasma to be incompressible, thermally inactive, and perfectly conducting \citep{kumar+2014phpl,kumar+2015phpl}. The set of dimensionless MHD equations is then given as: 

\begin{align}
 \frac{\partial{\bf{v}}}{\partial t} 
+ \left({\bf{v}}\cdot\nabla \right) {\bf{ v}} &=-\nabla p
+\left(\nabla\times{\bf{B}}\right) \times{\bf{B}}+\frac{\tau_a}{\tau_\nu}\nabla^2{\bf{v}},\label{stokes}\\  
\nabla\cdot{\bf{v}}&=0, \label{incompress1}\\
\frac{\partial{\bf{B}}}{\partial t}&=\nabla\times({\bf{v}}\times{\bf{B}}), \label{induction}\\
\nabla\cdot{\bf{B}}&=0, \label{solenoid} 
\end{align}

\noindent in usual notations. The magnetic field strength ${\bf{B}}$ and the plasma velocity ${\bf{v}}$ are normalised by the average magnetic field strength ($B_0$) and  the Alfv\'{e}n speed 
($v_a \equiv B_0/\sqrt{4\pi\rho_0}$ with $\rho_0$ representing the constant mass density), respectively. The plasma pressure $p$, the spatial-scale $L$, and the temporal scale $t$ are normalised by ${\rho {v_a}^2}$, the size of the system ($L_0$), and the Alfv\'{e}nic transit time ($\tau_a=L_0/v_a$), respectively. 
In Equation (\ref{stokes}), $\tau_\nu$ represents viscous diffusion time scale 
($\tau_\nu= L_0^2/\nu$), with $\nu$ being the kinematic viscosity.

 Notably, the incompressibility (Equation \ref{incompress1}) leads to the volume-preserving flow, an assumption routinely used in other works \citep{dahlburg+1991apj, aulanier+2005aa}. While compressibility is essential in exploring the thermodynamics of the coronal plasma \citep{ruderman&roberts2002apj}, in this work, our focus is on the changes in the magnetic topology of the initial quadrupolar magnetic configuration. Furthermore, utilizing the discretized incompressibility constraint, the pressure $p$ satisfies an elliptic boundary value problem on the discrete integral form of the momentum equation (Equation \ref{stokes}); cf. \citet{bhattacharyya+2010phpl} and the references therein. 

For the numerical solutions of the MHD equations, we use the well-established magnetohydrodynamic numerical model EULAG-MHD \citep{smolarkiewicz&charbonneau2013jcoph}. 
The model is an extension of the hydrodynamic model EULAG predominantly used in atmospheric and climate research \citep{prusa2008cf}. 
Here we discuss only the essential features of the EULAG-MHD code and refer the readers to \citet{smolarkiewicz&charbonneau2013jcoph} and references therein for detailed discussions.
The model is based on the spatiotemporally second-order accurate non-oscillatory forward-in-time multidimensional positive definite advection transport algorithm, MPDATA \citep{smolarkiewicz2006ijnmf}.  
Importantly, MPDATA has the proven dissipative property which, intermittently and adaptively, regularises the under-resolved scales by simulating magnetic reconnections 
and mimicking the action of explicit subgrid-scale turbulence models \citep{2006JTurb...7...15M} in the spirit of
Implicit Large Eddy Simulations (ILES)
\citep{grinstein2007book}. Such ILESs conducted with the model have already been successfully utilized to simulate reconnections to understand their role in the coronal dynamics \citep{prasad+2017apj,prasad+2018apj,2019ApJ...875...10N, kumar_2021SoPh, sanjay-2022}. In this work, the presented computation continues to rely on the effectiveness of ILES in regularizing the commencement of magnetic reconnections.

\section{Simulation Results}
The simulations are performed over a grid of $128 \times 128 \times 384$, in $x$, $y$, and $z$ directions. The magnetic field given by Equations (\ref{bx})-(\ref{bz}) (shown in Figure \ref{figure1}) serves as the initial magnetic field for the simulations. The initial velocity is set to zero (i.e., ${\bf{v}}=0$). 
The boundary condition along $z$ is kept open  by continuing the  ${\bf{B}}$ to the boundary in order to vanish the net magnetic flux passing through it {\citep{prasad+2018apj}}. Moreover, the boundaries along $x$ and $y$ are taken to be periodic. The simulations are done for two different viscosities $\nu=0.005$ and $0.01$.   
The values of the dimensionless constant $\tau_a / \tau_\nu$ in Equation (\ref{stokes}) are $\approx 10^{-4}$ and 
$\approx 10^{-3}$ for the viscosities $\nu=0.005$ and $0.01$ respectively, which are larger than its typical coronal value ($\approx 10^{-5}$) \citep{prasad+2018apj}. The higher values of  $\tau_a / \tau_\nu$ used in the simulations are only expected to speed up the dynamical evolution and reduce the computational cost without affecting the magnetic topology.  

\begin{figure}[htp]
\centering
\includegraphics[width=1\textwidth]{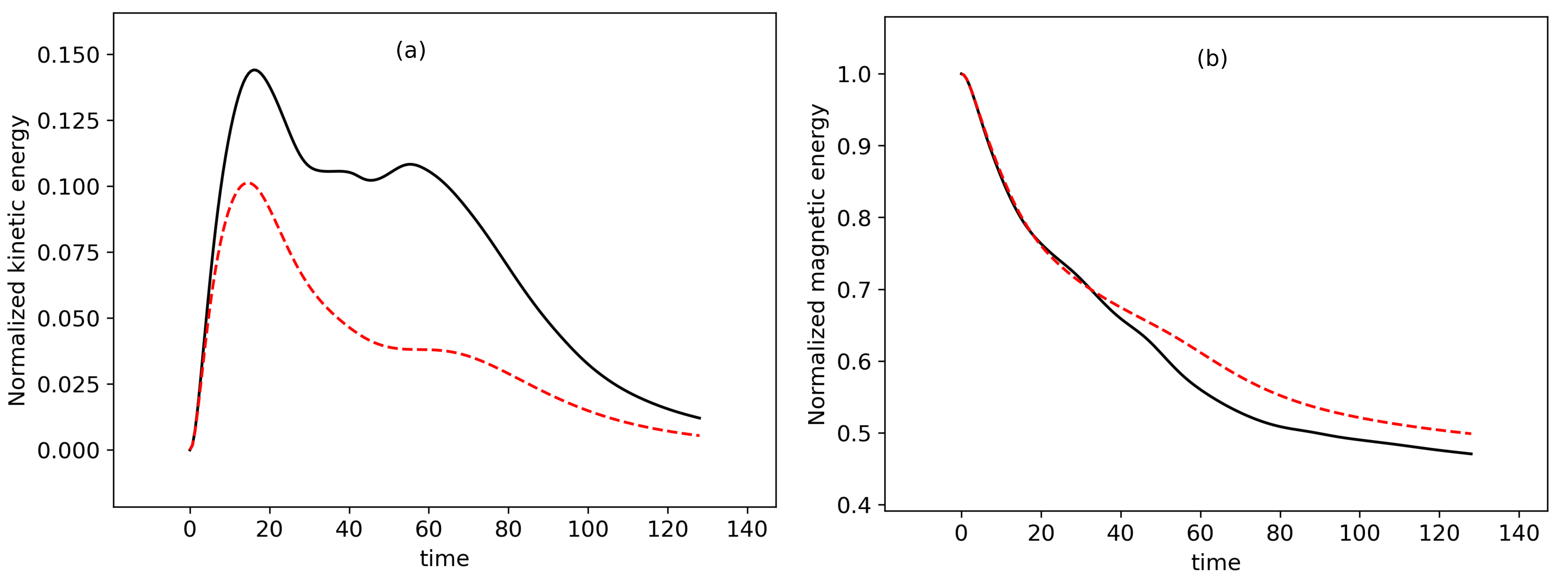}
\caption{  Time evolution of normalized kinetic and magnetic energies for  $\nu=0.005$ (black solid line) and $\nu=0.01$ (red dashed line). Time has units of seconds. The energies are normalized to the initial total energy. The plots highlight the formation of peaks in kinetic energy. Lower height in the kinetic energy peaks for higher viscosity is in agreement with the general understanding.  } \label{figure2a}
\end{figure}

For a general understanding of the simulated viscous relaxation, Figure \ref{figure2a} shows the time profile of the kinetic (panel (a)) and magnetic energies (panel (b)), normalized to the
initial total (magnetic + kinetic) energy. The black solid and red dashed lines correspond to viscosities $\nu=0.005$ and $0.01$, respectively. With an initially vanishing velocity field, the initial Lorentz force triggers the dynamical evolution and generates the plasma flow at the expense of the magnetic energy. The flow is then arrested by the viscous drag, which results in the formation of peaks in kinetic energy plots. Notably, the field lines being frozen into the plasma get deformed by the flow, modifying the initial distribution of the Lorentz force. The peak heights are lower for the simulation with $\nu=0.01$ in comparison to the simulation with $\nu=0.005$, which is expected due to the larger magnitude of the viscous drag force for $\nu=0.01$.

\begin{figure}[htp]
\centering
\includegraphics[width=0.75\textwidth]{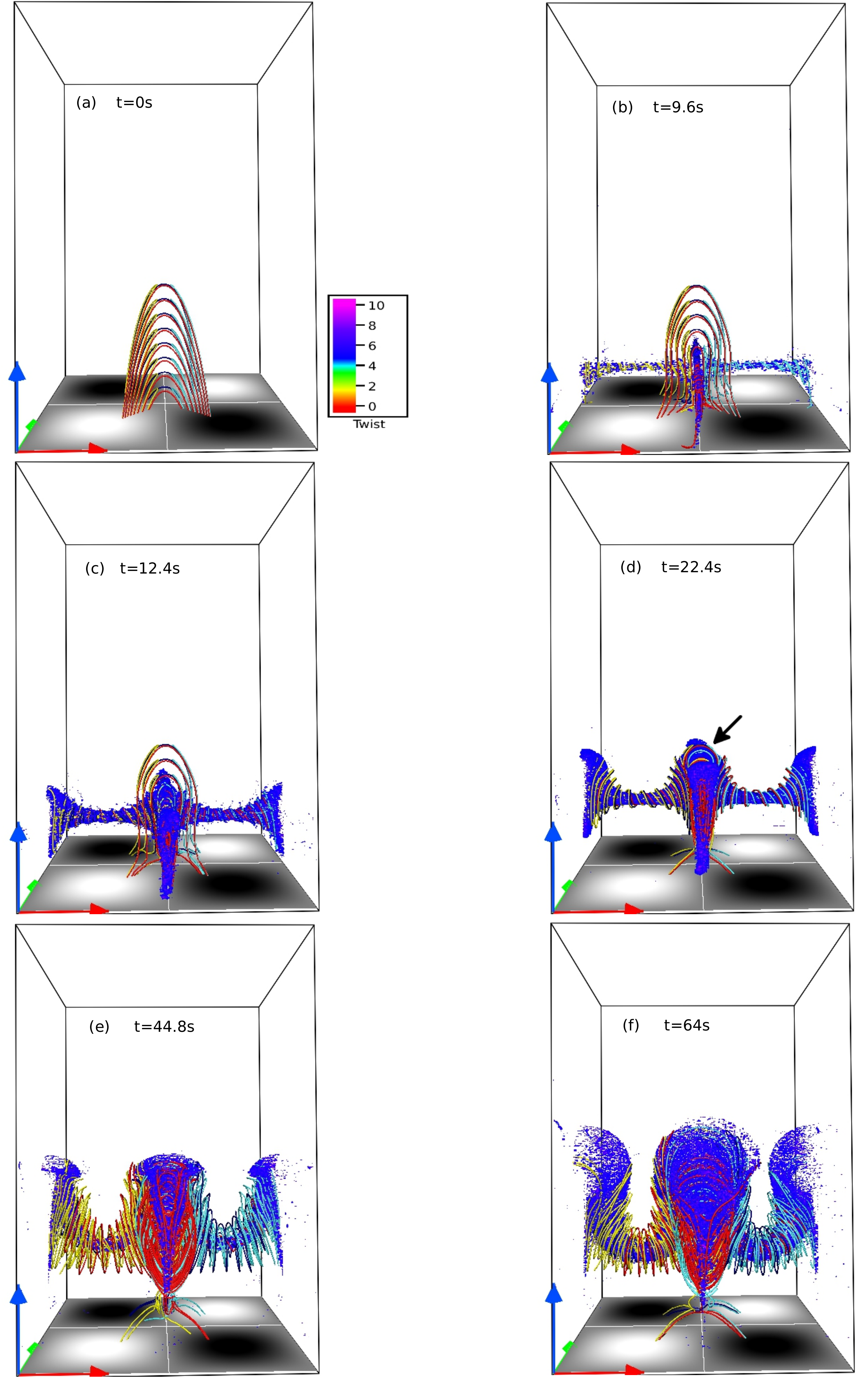}
\caption{ The MHD evolution of the initial quadrupolar magnetic configuration that is overlaid with the twist parameter. Initially, four separate MFRs are generated (panel (b)), which drift toward the X-type neutral lines at ($x, y, z$)=($\pi, \pi, z$), ($0, \pi, z$), ($\pi, 0, z$), ($\pi, 2\pi, z$), ($2\pi, \pi, z$) and fuse into complex magnetic structures (panels (c)-(e)). The structure around the X-type neutral line situated in the central part of the domain is marked by a black arrow in panel (d). The flux ropes also show non-uniform rising motion. Also, we have provided a supplementary animation of the figure for the high-cadence dynamics of the ropes.   } \label{figure2}
\end{figure}

\begin{figure}[htp]
\centering
\includegraphics[width=0.95\textwidth]{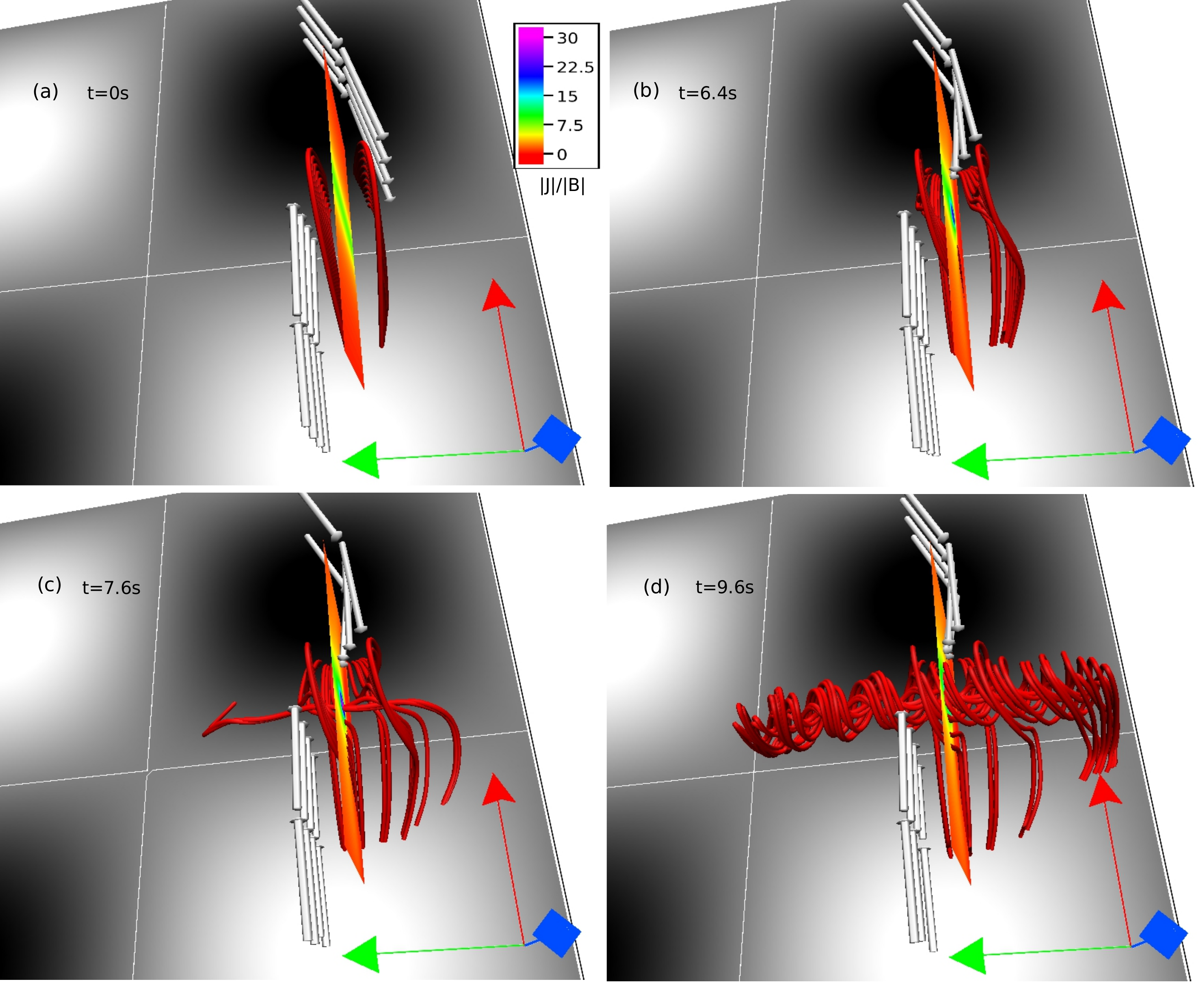}
\caption{Early phase of the dynamics of the two sets of the bipolar loops situated over the PIL between the P1 and N2 regions (Figure \ref{figure1}) overplotted with $|\mathbf{J}|/|\mathbf{B}|$ contours on a $y$-constant plane.  The Lorentz force is further shown with grey-colored arrows.  Notable are the contortions of the MFLs from the different loops under the favourable Lorentz force, which bring oppositely directed MFLs in proximity, as evident by the higher values of $|\mathbf{J}|/|\mathbf{B}|$ in panel (b)). This evolution leads to repeated reconnections and the formation of an MFR over the PIL (panels (c) and (d)).} \label{figure3}
\end{figure}

To explore the possibility of flux rope formation, in Figure \ref{figure2}, we first describe the evolution of the quadrupolar magnetic field configuration for the 
computation with viscosity $\nu=0.005$. The figure further shows the twist parameter, which is calculated by integrating field-aligned current ${\bf{J}} \cdot {\bf{B}}/B^2$ along a field line {\citep{Berger_2006, liu-2016}}. Figure \ref{figure2}(b) documents the generation of the twisted MFLs from the initial magnetic loops.  These field lines are co-located with the high values of the twist parameter --- suggesting the helical nature of the field lines.  Hence, the twisted field lines represent magnetic flux ropes located above the PILs. Initially, there are four separate MFRs, situated above the PILs of the different opposite polarities (P1, N1), (P1, N2), (P2, N1), (P2, N2); more clearly shown in Figure \ref{figure3}. With time, the magnetic flux of the MFRs increases, accompanied by their rise in the vertical direction (panels (c) and (d) of Figure \ref{figure2}). However, the rise of the MFRs is not uniform. Moreover, as the MFRs rise, they appear to interact with the pre-existing X-type neutral lines located at ($x, y, z$)=($\pi, \pi, z$), ($0, \pi, z$), ($\pi, 0, z$), ($\pi, 2\pi, z$), ($2\pi, \pi, z$) and, develop complex magnetic structures around them. The structure is more evident around the X-type neutral line that exists inside the computational domain at $(x, y, z)=(\pi, \pi, z)$ as shown in Figure \ref{figure1}(c) and marked by a black arrow in Figure \ref{figure2}(d). Further, the structures expand and rise with time (panels (e)-(f) of Figure \ref{figure2}). Similar evolution is also found for the computation with viscosity $\nu =0.01$ (not shown).

\begin{figure}[htp]
\centering
\includegraphics[width=0.9\textwidth]{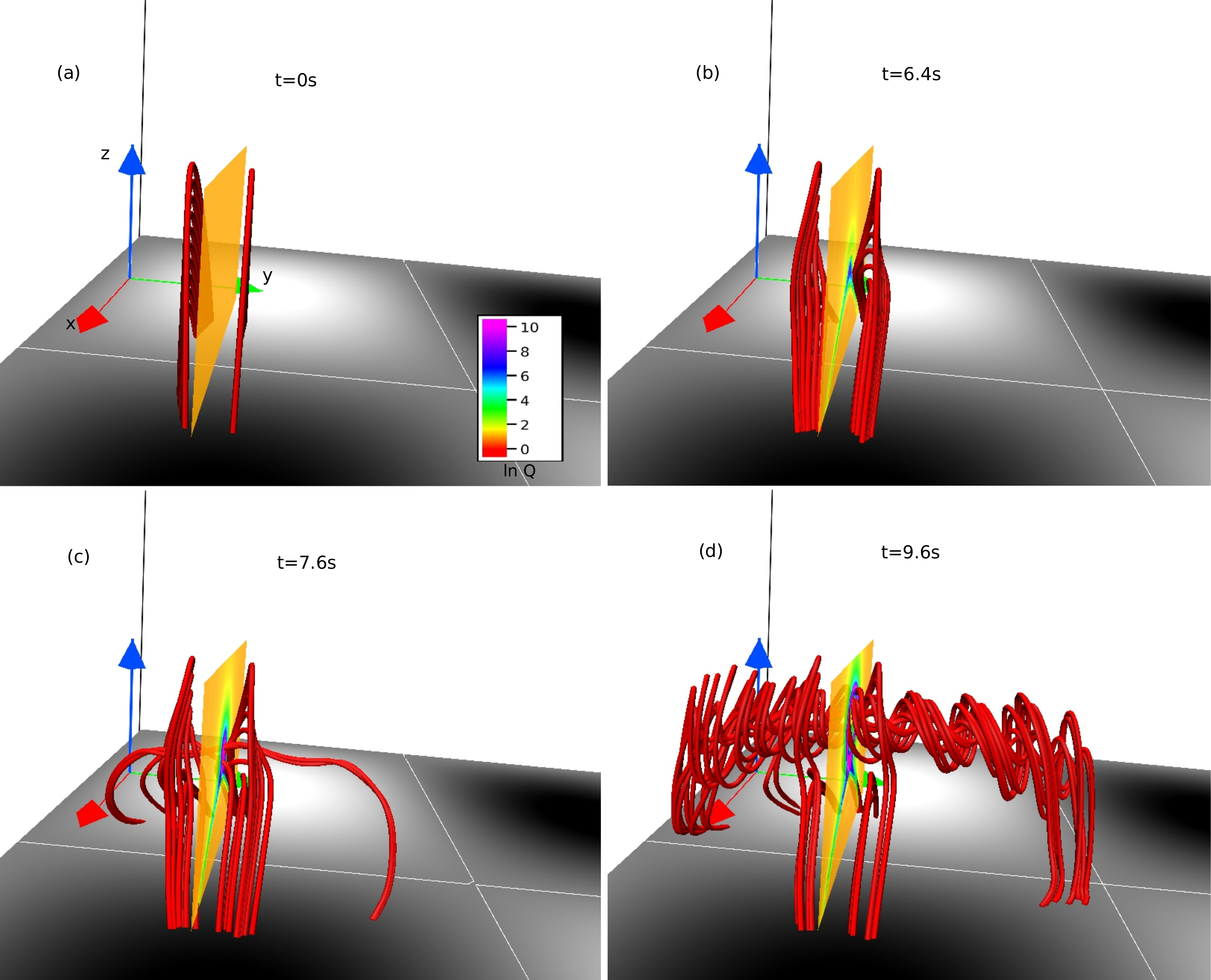}
\caption{Similar to Figure \ref{figure3}, but overplotted with $\ln Q$ contours on a $y$-constant plane. Higher values of $\ln Q$ in the vicinity of the contorted MFLs further confirm the generation of a high gradient in the magnetic field and consequent reconnections, which are responsible for the MFR formation.  } \label{figure4}
\end{figure}

To closely examine the mechanism of the MFR formation, in Figure \ref{figure3}, we focus on the early evolution of two sets of the bipolar magnetic loops located over the PIL corresponding to the regions P1 and N2 (Figure \ref{figure1}). The figure is further overlaid with the contours of $|\mathbf{J}|/|\mathbf{B}|$  on a $y$-constant plane and the Lorentz force (whose direction is  represented by the grey-colored arrows). The direction of the Lorentz force is such that the force pushes the two complementary anti-parallel field lines of the different set, located on the opposite sides of the PIL, toward each other (Figures. \ref{figure3}(a) and (b)). Consequently, the gradient in $\bf{B}$ sharply increases, as evident from a co-spatial enhancement in $|\mathbf{J}|/|\mathbf{B}|$. These dynamics lead to the onset of reconnections as the scales become under-resolved (Figure \ref{figure3}(c)). The reconnections continue in time and are central to developing an MFR over the PIL. 
In Figure \ref{figure4}, we overplot the two sets with the contours of the logarithm of squashing factor $Q$ on a $y$-constant plane to support the onset of the reconnections further. The $Q$-factor is a measure of the gradient in field line mapping of the magnetic field \citep{2002JGRA..107.1164T} and is calculated using the code of 
\citet{liu-2016}. Notable is the generation of high values of $\ln Q$ as the oppositely directed field lines approach each other. 
The high $\ln Q$ is co-spatial with the high $|\mathbf{J}|/|\mathbf{B}|$ (Figure \ref{figure3}) --- further supporting the development of the sharp gradient in $\bf{B}$ and consequent reconnections that lead to the MFR formation. After its formation, the legs of the MFR move towards the X-type neutral lines (panels (c) and (d) of Figures \ref{figure3} and \ref{figure4}). A similar process of reconnection also takes place over the other PILs located between (P1, N1), (P2, N1), and (P2, N2)-regions and is responsible for the formation of MFRs over the PILs (not shown).

\begin{figure}[htp]
\centering
\includegraphics[width=0.9\textwidth]{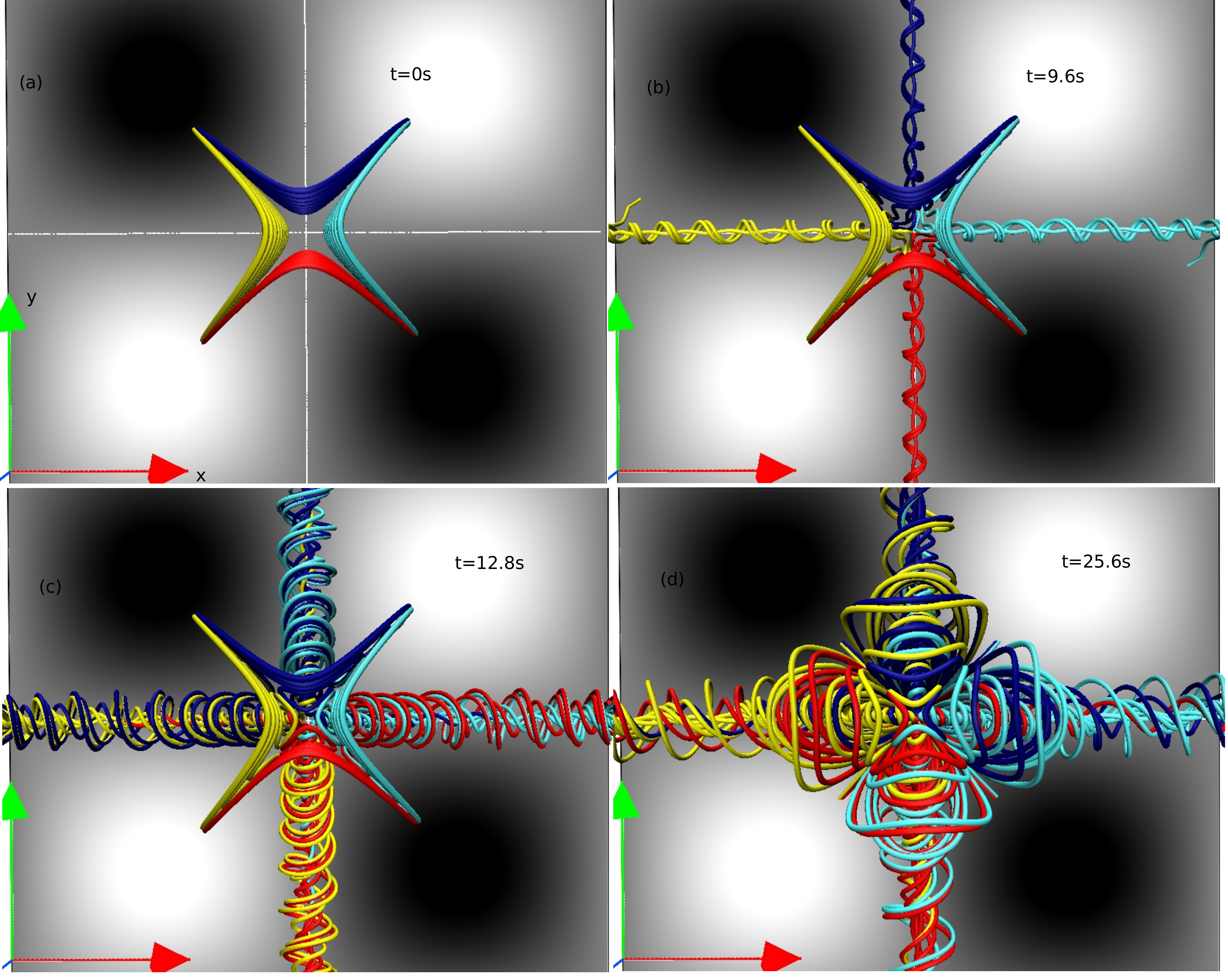}
\caption{The zoomed-in top view of the MHD evolution of the magnetic field lines near the X-type neutral line inside the computational domain. Reconnections of the flux ropes at the X-type neutral line are evident --- leading to the formation of complex magnetic structures.  } \label{figure5}
\end{figure}

In Figure \ref{figure5}, we plot the top view of the time evolution of the magnetic field lines near the X-type neutral line located inside the central region of the domain at $(x, y, z)=(\pi, \pi, z)$. The figure illustrates that, as the legs of the four MFRs approach the X-type neutral line, they start to reconnect at the neutral line and develop the complex magnetic structure around the neutral line. As these reconnections repeat, more and more magnetic flux is sucked into the central region --- shaping the structure and, ultimately, expanding it. Formation of the complex structures through similar dynamical evolution is also observed around the other neutral lines situated at the boundaries of the domain with the coordinates $(x, y, z)$=($0, \pi, z$), ($\pi, 0, z$), ($\pi, 2\pi, z$), ($2\pi, \pi, z$) (not shown).     

\begin{figure}[htp]
\centering
\includegraphics[width=0.9\textwidth]{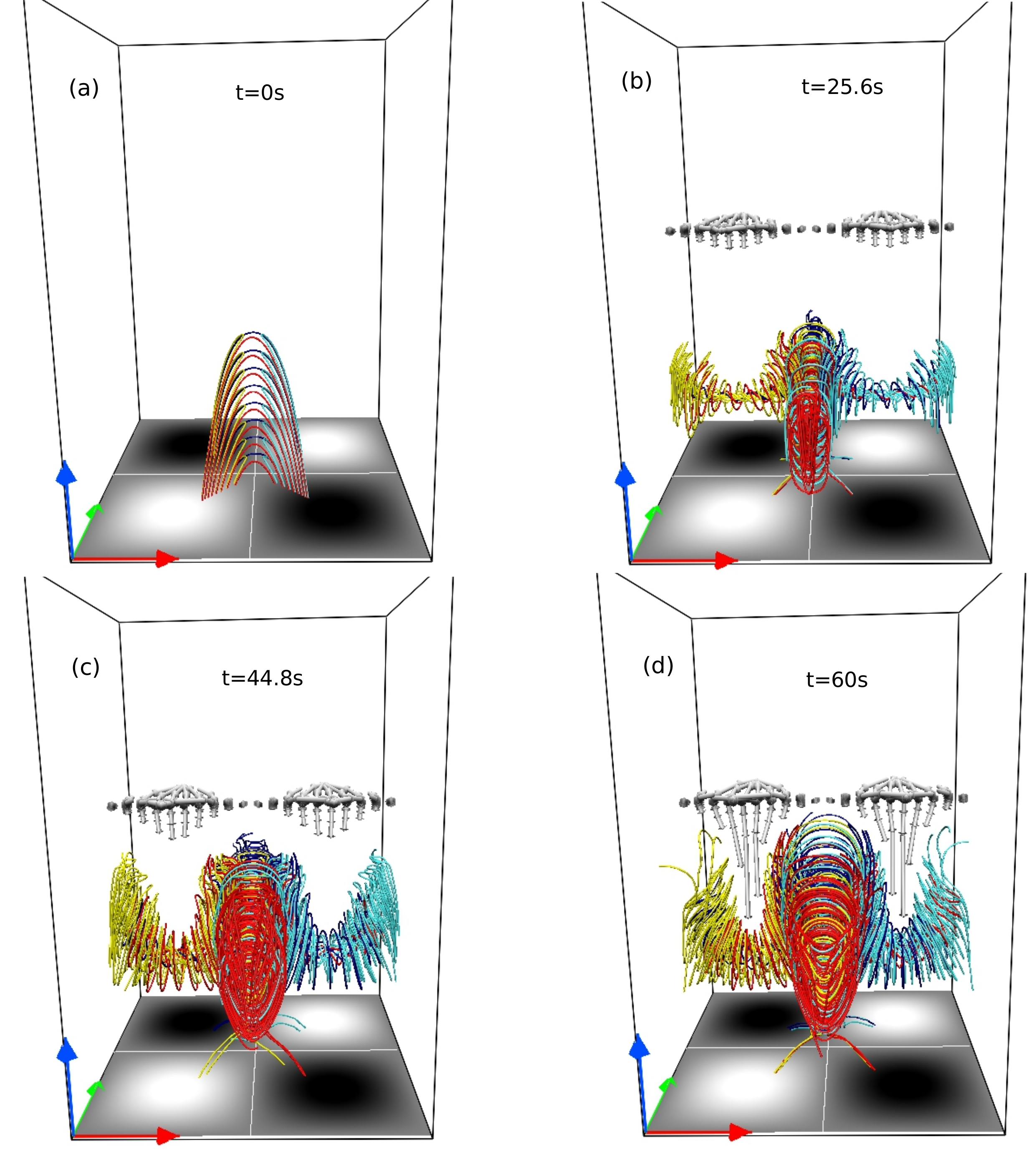}
\caption{Evolution of the quadrupolar MFLs overlaid with the Lorentz force (grey-colored arrows). The almost negligibly-small Lorentz force over the complex magnetic structures is notable, while the force is mostly vertically downward over the remaining parts of the ropes. Hence, this allows the faster rise and expansion of the structures than the rest.  Also, a supplementary animation is provided for a high-cadence evolution.  } \label{figure6}
\end{figure}

Further, the outflows generated from the repeated reconnections occurring below the flux ropes push them in the vertical direction and lead to the ascent of the flux ropes. Notable is the non-uniform ascent of the flux ropes. 
The flux ropes show a faster rise and expansion near the neutral lines than away from them (see Figure \ref{figure2}). To explore this non-uniform ascent, in Figure \ref{figure6}, we plot the evolution of the flux ropes overlaid with the Lorentz force (represented by the grey-colored arrows). The direction and length of the arrows mark the direction and magnitude of the Lorentz force. The figure shows that the Lorentz force is almost negligible around the X-type neutral line (over the complex magnetic structure), while the force is vertically downward away from the neutral line. 
 Moreover, the magnitude of the vertically downward force appears to increase with time. 
As a result, the overlying Lorentz force does not affect the rise of the flux ropes in the vicinity of the neutral line, while the vertically downward force restricts the rise of the ropes away from the neutral line. This result suggests that the non-uniform rise of the ropes is related to the non-uniform distribution of the overlying Lorentz force.

\begin{figure}[htp]
\centering
\includegraphics[width=0.95\textwidth]{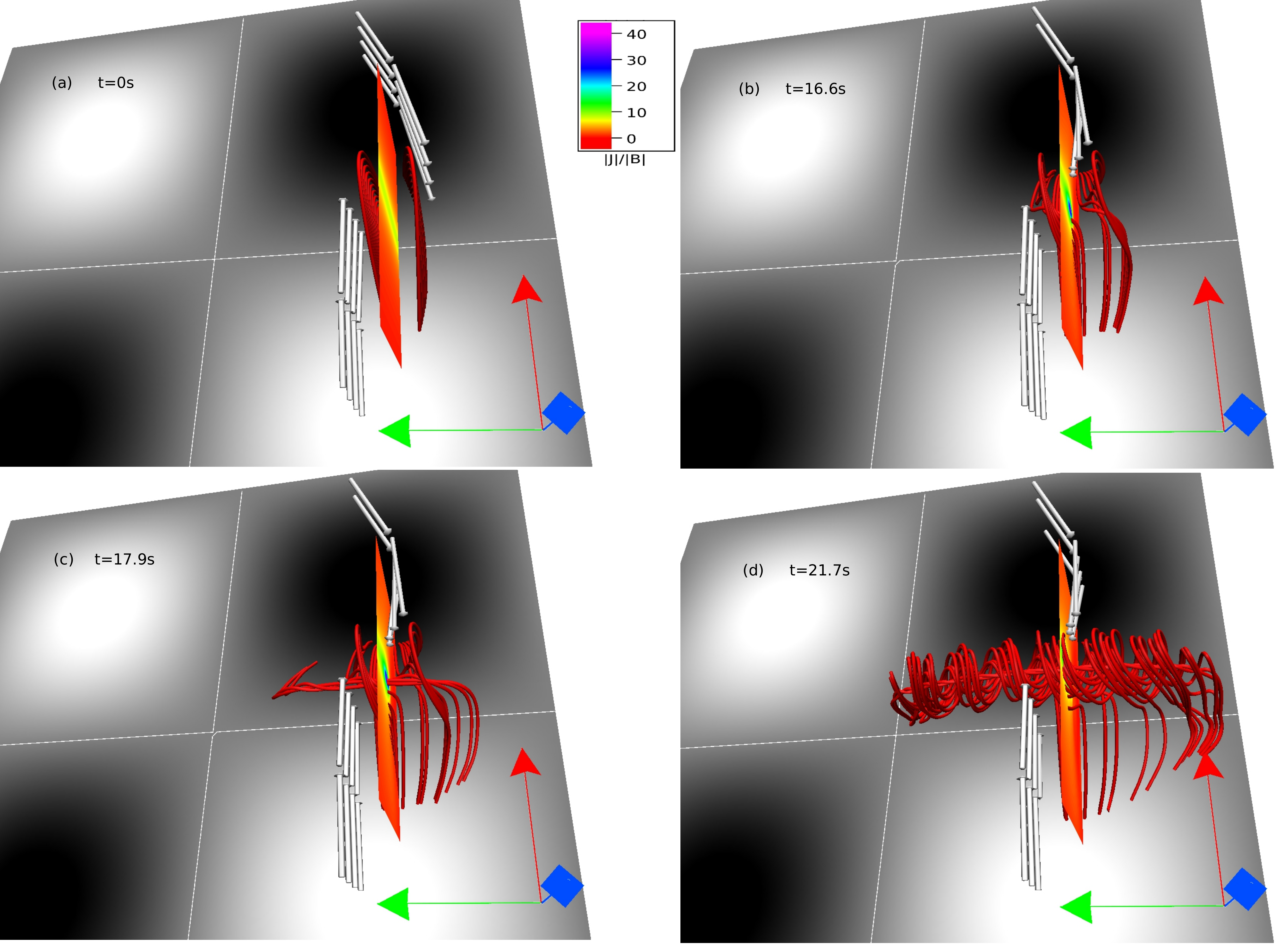}
\caption{Similar to Figure \ref{figure3}, but with the higher resolution of $160 \times 160 \times 480$. 
 Notable is the higher value of $|\mathbf{J}|/|\mathbf{B}|$ in compare to Figure \ref{figure3}. The field line evolution is morphologically identical to the ones shown in Figure \ref{figure3}.} \label{figure7}
\end{figure}

To show the convergence of the simulation results with increasing resolution, we have also performed an additional higher resolution simulation over a grid of $160 \times 160 \times 480$, in $x$, $y$, and $z$ directions with viscosity $\nu=0.01$. Figure \ref{figure7} illustrates the time evolution of the two sets of the bipolar magnetic loops in the same sub-domain as shown in Figure \ref{figure3}. Moreover, similar to Figure \ref{figure3}, Figure \ref{figure7} is also overplotted with the contours of $|\mathbf{J}|/|\mathbf{B}|$  on a $y$-constant plane and the Lorentz force (whose direction is marked by the grey-colored arrows). For the higher resolution simulation, the maximum value of $|\mathbf{J}|/|\mathbf{B}|$ is increased (see panel (a) of Figures \ref{figure3} and \ref{figure7}) --- agreeing with the general understanding that the gradient in the magnetic field  enhances with an increase in grid resolution. The field line evolution in Figure \ref{figure7} is geometrically identical to the ones depicted in Figure \ref{figure3} --- confirming the convergence of the results with increasing resolution. Moreover, Figure \ref{figure7} documents a delay in the onset of the reconnection, with an increase in grid resolution. The delay is expected as the under-resolved scales develop later in time with an increased resolution.

\section{Summary}
The presented MHD simulations explore the magnetic flux rope formation process in the presence of a quadrupolar magnetic field  topology. The initial non-force-free magnetic field is constructed analytically by modifying a three-dimensional linear force-free field having field line geometry similar to the observed coronal loops. The configuration consists of two positive polarity regions, P1 and P2, and two negative polarity regions, N1 and N2. Because of these different polarity regions, the initial field has an X-type neutral line inside the computational domain. In addition, X-type neutral lines also reside at the domain's boundaries due to the periodicity in the lateral directions. Furthermore, the field supports Lorentz force, which naturally generates the simulated dynamics from an initial motionless state. The plasma evolution is idealized to be viscid, incompressible, and thermally homogeneous. Furthermore, the locally adaptive dissipation of the MPDATA scheme mimics the magnetic reconnections in response to the development of the under-resolved scales.  

In the simulations, we first notice the movement of oppositely directed  field lines of bipolar loops located above the PILs (separating the different polarity regions) towards each other. This evolution leads to a steep enhancement in the magnetic field gradient and the generation of under-resolved scales. Consequently, repetitive reconnections initiate and account for the formation of flux ropes over the PILs. In addition, the outflow generated by these reconnections pushes the flux ropes in the upward direction and contributes to their ascent. 

 With time, the legs of the magnetic flux ropes move toward the X-type neutral lines and start reconnecting at the line --- leading to the generation of complex magnetic structures around the neutral lines. Hence, the simulations demonstrate the topologically complex MFR evolution in the quadrupolar configuration than those found in bipolar magnetic loops {\citep{sanjay2016}},  which is a key finding of the paper. 
Furthermore, the rise of the flux ropes is found to be non-uniform. The flux rope near the neutral lines exhibits a faster rise than away from the neutral lines. The non-uniform rise is attributed to the non-uniform generation of the overlying Lorentz force. The Lorentz force is negligible near the neutral lines and enhances with a vertically downward direction as the distance from the neutral lines increases.

Overall, the reported simulations identify spontaneous
repeated magnetic reconnections as the initial driver for the flux rope formation and triggering its ascent in the quadrupolar magnetic configuration. Notably, the ascent of the flux ropes having underlying reconnections is in harmony with contemporary observations at multiple channels (hard X-ray and extreme ultraviolet) that reveal intense localized brightenings below a rising flux rope {\citep{2011ApJ...732L..25C, 2015ApJ...807..101K}}. More importantly, the pre-existing X-type neutral line and the newly formed flux ropes find morphological similarity to the brightenings observed in extreme ultraviolet images during the flaring events in the quadrupolar magnetic configurations (cf. \citet{2017ApJ...842..106K, 2022ApJ...926..143M}).
The similarity indicates that the pre-existing X-type neutral points as well as the formation of the flux ropes in the magnetic configurations can play a crucial role in initiating the flaring events.  
However, the presented simulations, idealized with an analytically constructed initial field and the assumed incompressibility, are only partly conclusive. 
Therefore, we set a future goal to perform a fully compressible MHD simulation initiated by the extrapolated coronal magnetic field with the observed quadrupolar magnetic configuration, which is expected to provide more realistic coronal dynamics and can be directly compared with solar observations. 

\subsection*{Data availability statement}
The datasets generated for this study are available on request to the corresponding author.

\subsection*{Acknowledgments}
We acknowledge the visualization software VAPOR (www.vapor.ucar.edu) for generating relevant graphics.
AP would also like to acknowledge the support of the Research Council of Norway through its Centres of Excellence scheme, project number 262622, and Synergy Grant number 810218 459 (ERC-2018-SyG) of the European Research Council. AP also acknowledges partial support from NSF award AGS-2020703. SSN acknowledges the NSF-AGS-1954503 and NASA-LWS-80NSSC21K0003 grants.

\bibliographystyle{jphysicsB}
\bibliography{refs.bib}

\end{document}